\documentclass[reprint,%
]{revtex4-1}

\usepackage{natbib}
\setcitestyle{authoryear,round}
\usepackage[utf8]{inputenc}

\usepackage{braket}
\usepackage{float}      
\usepackage{bbold}      
\usepackage[caption=false]{subfig}
\usepackage{subfig}

\usepackage{bm}         
\usepackage{color}
\usepackage{nicefrac}   
\usepackage{setspace}   
\usepackage{booktabs}   
\usepackage{multirow}
\usepackage{umoline}
\usepackage{amsmath}
\usepackage{amsfonts}
\usepackage{dcolumn}
\usepackage{bm}
\usepackage{graphicx}

\begin{document}

\title{Quantum Hall and Synthetic Magnetic-Field Effects in Ultra-Cold Atomic Systems  \\
\vspace{0.1cm}
\begin{minipage}{0.7\textwidth}
\centering
\textmd{\small{{\it Invited Contribution to} Encyclopedia of Condensed Matter Physics, {\it 2nd edition}}}
\end{minipage}
\vspace{-0.1cm}
}

\author{Philipp Hauke}
\email{philipp.hauke@unitn.it}
\affiliation{INO-CNR BEC Center and Department of Physics, University of Trento, Via Sommarive 14, I-38123 Trento, Italy}
\affiliation{INFN-TIFPA, Trento Institute for Fundamental Physics and Applications, Trento, Italy}

\author{Iacopo Carusotto}
\email{iacopo.carusotto@unitn.it}
\affiliation{INO-CNR BEC Center and Department of Physics, University of Trento, Via Sommarive 14, I-38123 Trento, Italy}

\date{\today}

\begin{abstract}
    In this Chapter, we give a brief review of the state of the art of theoretical and experimental studies of synthetic magnetic fields and quantum Hall effects in ultracold atomic gases. We focus on integer, spin, and fractional Hall effects, indicate connections to topological matter, and discuss prospects for the realization of full-fledged gauge field theories where the synthetic magnetic field has its own dynamics. The advantages of these systems over traditional electronic systems are highlighted. Finally, interdisciplinary comparisons with other synthetic matter platforms based on photonic and trapped-ion systems are drawn. We hope this chapter to illustrate the exciting progress that the field has experienced in recent years.  
\end{abstract}

\maketitle

\section{Key points/Objectives}
This chapter aims to 
\begin{itemize}
    \item illustrate how synthetic magnetic fields can effectively be generated in neutral matter;
    \item highlight breakthroughs in realizing integer and spin quantum Hall effect; 
    \item showcase steps towards strongly interacting systems displaying the fractional quantum Hall effect; 
    \item illustrate some of the rich physics that can be achieved when the atoms act back on the synthetic gauge fields; 
    \item emphasize the fascinating progress of the field and the outstanding main challenges.
\end{itemize}

\section{Introduction}
\label{sec:introduction}

The discovery of the quantum Hall (QH) effects in two-dimensional electron gases subject to a strong magnetic field has revolutionized our view of quantum condensed-matter physics and has opened a new field of research at the crossroad with quantum information and topology~\citep{Tong:QHbook}.
While usual phase transitions such as ferromagnetism or even Bose--Einstein condensation are described within the Landau--Ginzburg theory by the onset of a macroscopic order parameter, topological states of matter cannot be revealed by the measurement of any local observable and instead are characterized by subtle quantum correlations between the constituent particles~\citep{wen1995topological,Kitaev2006,levin2006detecting,Wen2013}. 
The most direct signature of integer and fractional quantum Hall states is the quantized value of the transverse conductivity in units of $e^2/h$. Besides opening windows into novel  physical phenomena of fundamental interest, such states have been also anticipated for practical applications in quantum information processing, e.g., as a promising platform to transmit quantum information and protect it  from decoherence~\citep{Nayak:RMP2008}.

Beyond the traditional observation of quantum Hall effects in electron gases in solid-state devices, a strong activity is presently being devoted to synthetic quantum matter systems~\citep{ozawaprice2019topological} such as gases of ultracold atoms~\citep{pitaevskii2016bose,cooper2008rapidly}, spin and phonon systems realized in trapped ions~\citep{Bermudez2011,nigg2014,Manovitz2020,Geier2021}, or fluids of strongly interacting photons in nonlinear topological photonics devices~\citep{carusotto2013quantum,carusotto2020photonic}. In all these systems, the charge neutrality of the constituent particles requires introducing a so-called synthetic magnetic field in order to observe QH physics~\citep{dalibard2011colloquium,goldman2014light,dalibard2015introduction}; at the same time, such systems allow for a broader range of techniques for the manipulation and diagnostics well beyond the standard transport and optical probes of solid state systems~\citep{yoshioka2002quantum}, which opens exciting new perspectives to experimental studies. While a number of phenomena related to the integer quantum Hall (IQH) effect have been observed in the last years for non-interacting particles in suitably designed band structures for either atoms or photons, fractional quantum Hall (FQH) effects require strong interactions among particles and low temperatures, and are still a subject of intense work.

In what follows, we give a timely account of the state of the art of experimental and theoretical research towards the realization of FQH liquids of ultracold atoms. 
We also discuss relatively novel directions, where the magnetic vector potential acquires its own dynamics, leading to interesting backaction effects of matter particles onto the synthetic magnetic field and even to full-fledged gauge theories reminiscent of quantum electrodynamics. 
This chapter thus complements other reviews on synthetic gauge fields~\citep{dalibard2011colloquium,goldman2014light,dalibard2015introduction}, on rotating atomic clouds~\citep{cooper2008rapidly}, and on topological effects in synthetic quantum matter \citep{Zhang2018,ozawaprice2019topological,Cooper2019}. 
While the main focus of our discussion here will be on ultracold atoms, connections to other promising platforms such as trapped ions or photonic devices will also be drawn.

This review chapter is organized as follows. 
In Sec.~\ref{sec:Synthetic_magnetic_fields}, we recapitulate approaches to imbuing ultracold atomic gases with synthetic magnetic fields, a key prerequisite for observing QH physics. Sections~\ref{sec:IQH} and~\ref{sec:FQH} discuss recent advances in cold atom systems in observing the IQH and FQH, respectively. 
In Sec.~\ref{sec:LGTs}, we highlight some recent developments in inducing backaction of the matter system onto the synthetic magnetic field, leading to the generation of dynamical Peierls phases and even to pioneering steps in the creation of lattice gauge theories. 
Section~\ref{sec:Conclusion} contains our conclusions and some further outlooks.

\section{Synthetic magnetic fields}
\label{sec:Synthetic_magnetic_fields}

The motion of quantum mechanical particles of mass $m$ and charge $q$ in a classical background magnetic field is described  via the minimal coupling Hamiltonian
\begin{equation}
    \label{eq:Hkincont}
    H_\mathrm{kin}=\frac{(\mathbf{P}-\frac{q}{c}\mathbf{A})^2}{2m}\,
\end{equation}
in terms of the vector potential $\mathbf{A}(\mathbf{r})$, related to the magnetic field by $\mathbf{B}=\textrm{rot}\mathbf{A}$. 
In a discrete, spatially-periodic lattice geometry, this Hamiltonian can be reformulated in terms of a non-trivial hopping phase in the tight-binding Hamiltonian,
\begin{equation}
    \label{eq:Hkinlatt}
    H_\mathrm{kin}=-J \sum_{\braket{i,j}} \psi_i^\dagger U_{i,j} \psi_j \,,
\end{equation}
where $J$ is the bare hopping amplitude, $\psi_i^\dagger$ and $\psi_i$ are fermionic or bosonic creation and annihilation operators at lattice site $i$, respectively, and 
$U_{i,j}=\mathrm{exp}(i \frac{q}{c} \int_{\mathbf{r}_i}^{\mathbf{r}_j} \mathrm{d}\mathbf{r}\cdot \mathbf{A}(\mathbf{r}))$
is the Peierls phase, corresponding to the Aharonov--Bohm phase accumulated by the particle when moving from site $i$ to site $j$. Starting from these formulations, it is straightforward to derive the celebrated features of quantum mechanical motion in magnetic fields, such as cyclotron motion and Landau levels~\citep{Tong:QHbook,Girvin}.

Since atoms are charge neutral ($q=0$), it is impossible to observe this physics using standard magnetic fields, whose effect on atoms typically reduces to Zeeman shifts of the internal levels. More subtle manipulations are therefore necessary to generate effective vector potentials affecting the orbital atomic motion along the lines of Eqs.~\eqref{eq:Hkincont} or \eqref{eq:Hkinlatt}.
Over the years, a number of strategies to generate such {\em synthetic magnetic fields} have been developed. Here, we will briefly review the most promising and/or physically transparent ones~\citep{dalibard2011colloquium,goldman2014light,dalibard2015introduction}. 
As the basic object realized in the experiments is typically the vector potential, gauge-invariance of the framework raises interesting subtleties in the conceptual analysis of the experiments~\citep{LeBlanc_2015}. 
Though synthetic gauge fields give rise also to other intriguing effects, e.g., supersolid behavior~\citep{Li2017,Putra2020,Geier2021Exciting}, our attention in this review will be focused on the family of integer and fractional Quantum Hall effects.

\begin{figure}
    \centering
    \includegraphics[width=0.9\columnwidth]{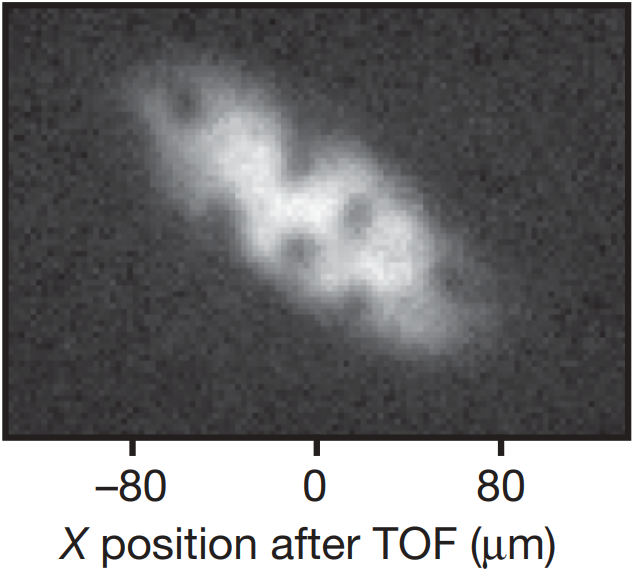}
    \caption{Quantized vortices nucleated in an atomic BEC under the effect of a synthetic magnetic field. Figure adapted from~\citep{lin2009synthetic}.}
    \label{fig:vortices}
\end{figure}
Historically, the first strategy was based on the mathematical analogy between the Coriolis force $\mathbf{F}=\mathbf{v}\times \mathbf{\Omega}$ in a rotating frame and the Lorentz force $\mathbf{F}=q\,\frac{\mathbf{v}}{c}\times \mathbf{B}$ in a magnetic field. By putting the atom trap into mechanical rotation around some axis, the atomic cloud gets to equilibrate in a rotating frame where the angular speed plays the role of the magnetic field. Over the years, this approach has led to pioneering advances for weakly interacting Bose--Einstein condensates, such as the observation of periodic arrays of quantized vortices~\citep{bretin2004fast,abo2001observation,schweikhard2004rapidly} resembling the Abrikosov vortex arrays generated by magnetic fields in type-II superconductors. 

A conceptually different approach to generate a synthetic magnetic field is based on the Berry phase that moving atoms experience when they are dressed by a suitable combination of optical, micro- and radio-wave, and static electromagnetic fields~\citep{dum1996gauge,Juzeliunas2004,Ruseckas2005}: in this framework, the roles of the vector potential and of the magnetic field are played by the Berry connection and the Berry curvature, respectively. Remarkable outcomes of this approach include the nucleation of quantized vortices under the effect of strong enough synthetic fields~\citep{lin2009synthetic}, as highlighted in Fig.~\ref{fig:vortices}. 

Discrete, spatially periodic geometries realized by imposing optical lattice potentials offer additional tools to generate synthetic magnetic fields via non-trivial hopping phases.  
Complex hopping phases induced by inter-site Raman processes were first proposed in Ref.~\citep{jaksch2003creation,Mueller2004,Sorensen2005,Osterloh2005} and then implemented to realize, e.g., an atomic Harper--Hofstadter model~\citep{miyake2013realizing,aidelsburger2013realization}.

Alternatively, non-trivial band geometries and topologies have been realized in a Floquet framework based on periodic shaking or time-modulation of the lattice potential \citep{Eckardt2015,Bukov2015}. 
By suitably choosing the shaking protocol, e.g., such that the periodic shaking function explicitly breaks time-reversal symmetry, one can obtain---in a time-averaged or stroboscopic Floquet description---an effective Hamiltonian that has a non-trivial Peierls phase \citep{Oka2009,Kitagawa2010,Kitagawa2011,Kolovsky2011,HaukeEckardt,Hauke2012,goldman2014periodically}. 
This approach has led to the realization of the Haldane~\citep{jotzu2014experimental} and the Harper--Hofstadter model \citep{aidelsburger2015measuring}. 

An exciting new frontier of these investigations is to combine the internal and orbital dynamics of atoms into a single geometric entity so to realize configurations with a larger number of effective spatial dimensions than the usual three spatial ones~\citep{celi2014synthetic}, where to observe new magnetic and topological effects that are peculiar of high dimensionalities~\citep{price2015four}. Pioneering steps in the direction of realizing such {\em synthetic dimensions} were reported in the experiments of Refs.~\citep{mancini2015observation,stuhl2015visualizing}.

\section{Integer Quantum Hall effect}
\label{sec:IQH}

\begin{figure}
    \centering
    \includegraphics[width=0.99\columnwidth]{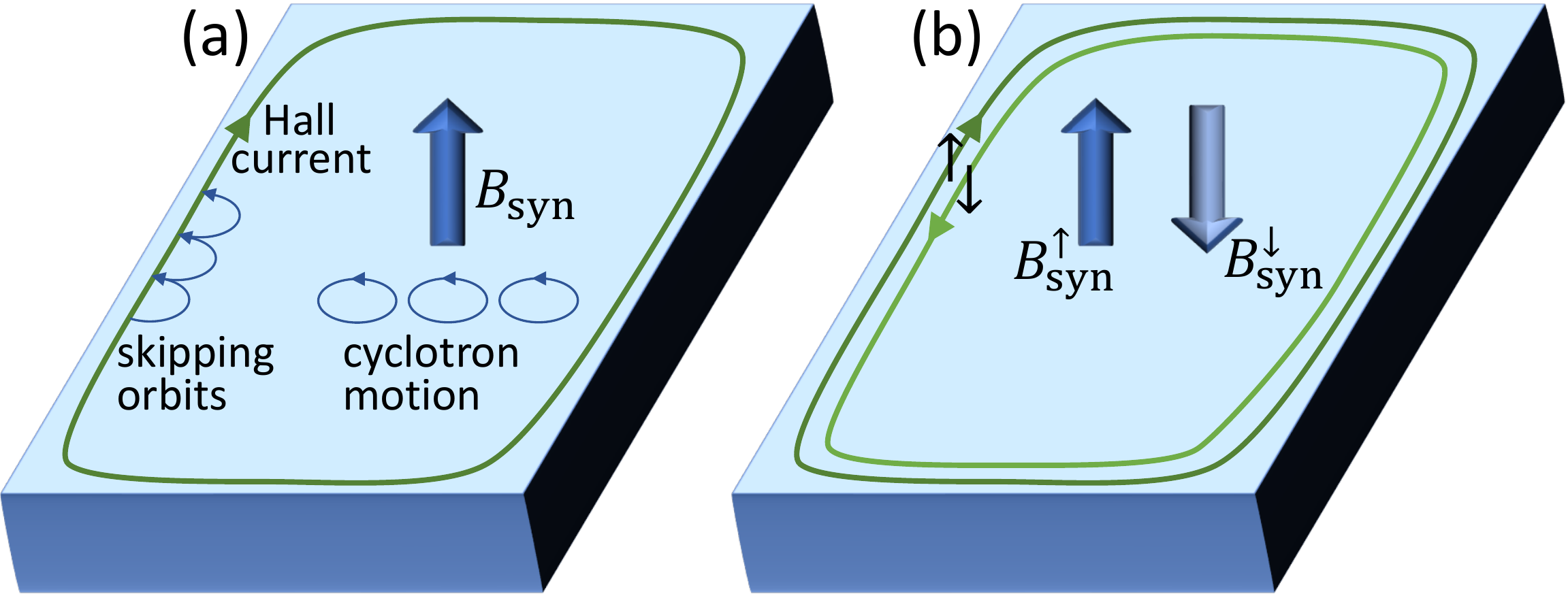}
    \caption{(a) Integer Quantum Hall effect. A synthetic magnetic field pierces a cloud of ultracold neutral atoms, induced, e.g., by trap rotation, Berry phase imprinting, or lattice shaking. Within the bulk, this leads to cyclotron motion. The corresponding skipping orbits at the boundary add up to a chiral edge current, which is protected by the topology of the sample.  
    (b) Spin Hall effect. Two (pseudo-)spin states (e.g., two atomic hyperfine states) see an opposite synthetic magnetic field. As a result, the system has vanishing net charge transport, but the presence of helical edge states results in a topologically protected spin current around its boundaries.
    }
    \label{fig:IQHE-spinHE}
\end{figure}

In electronics, the main signature of the Integer Quantum Hall (IQH) effect consists in a quantized value of the transverse conductance to integer multiples of $e^2/h$ whenever electrons completely fill an integer number of Landau levels~\citep{Tong:QHbook}. Soon after the discovery of the IQH, it was realized that the quantized conductance can be related to integer-valued topological invariants characterizing the geometry of the electronic bands~\citep{thouless1982quantized}. 
In a modern understanding, IQH systems are then interpreted as an example of the wider class of topological insulators, where the bulk-boundary correspondence translates topological properties of the bands of the insulating bulk into chiral currents around the edges of the sample \citep{hasan2010colloquium}, see Fig.~\ref{fig:IQHE-spinHE}(a). 
Since these features are related to integer-valued topological invariants, they are resilient against disorder and similar types of small perturbations.

Based on such insights, the IQH effect could be studied  
in a number of atomic systems: even though conductivity experiments analogous to electronic ones are not naturally performed in atomic systems, many other observables are available to obtain an even deeper information on the properties of this state of matter.

\subsection{IQH in cold atoms}

Anomalous Hall effects were first predicted in the atomic context in Ref.~\citep{dudarev2004spin} and experimentally observed in honeycomb-like lattices in Ref.~\citep{jotzu2014experimental} as a lateral displacement of the atoms whenever their $k$-space distribution is pushed across a finite-Berry-curvature region. A quantized displacement analogous to the IQH was then observed for an atomic cloud uniformly filling a band~\citep{aidelsburger2015measuring}. 
A precursor of this was the measurement of the Zak phase characterizing topological Bloch bands in a one-dimensional optical lattice \citep{Atala2012}. 

\begin{figure}
    \centering
    \includegraphics[width=0.9\columnwidth]{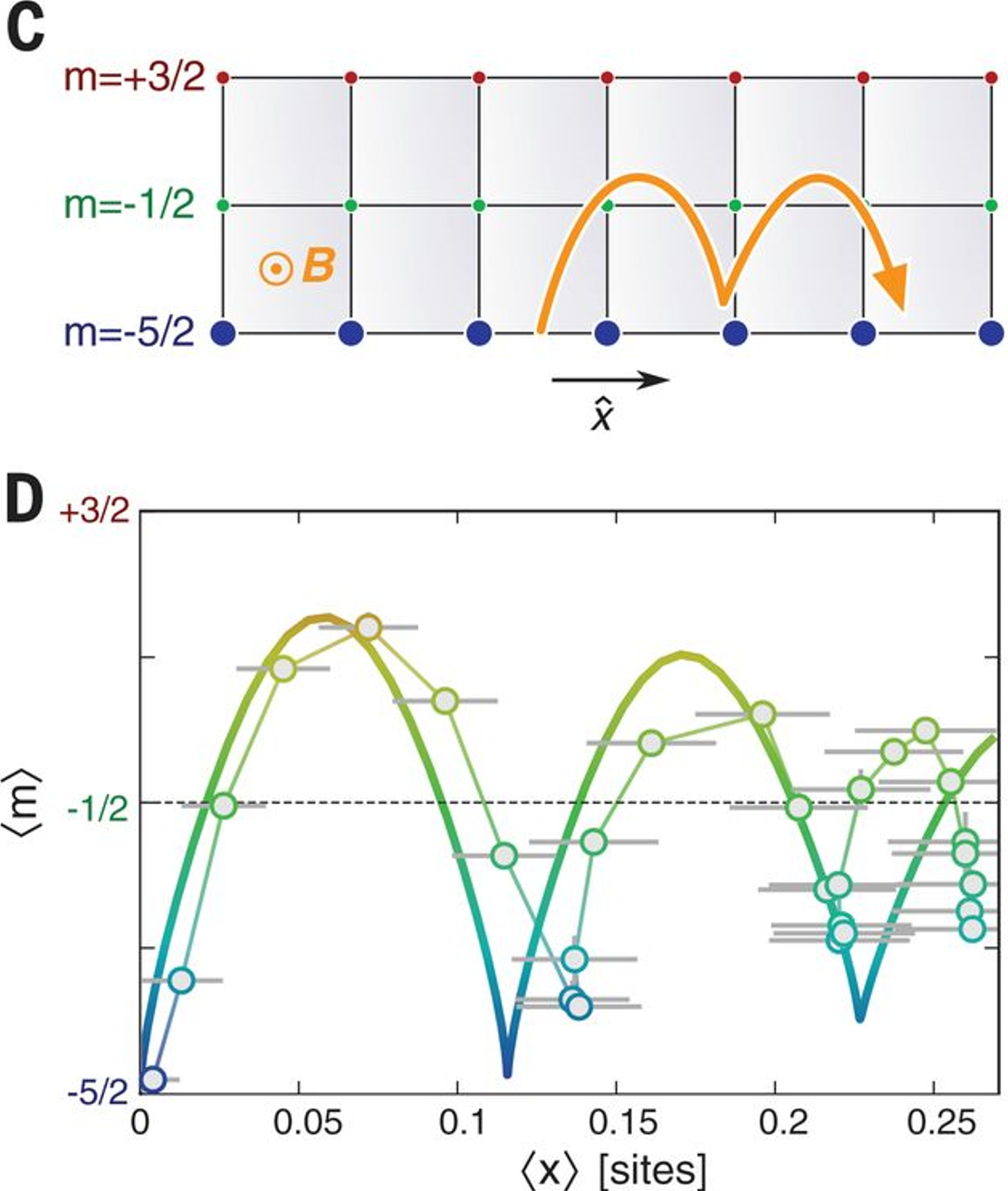}
    \caption{Experimental observation of edge-cyclotron orbits in a synthetic three-leg ladder system based on a spatial $x$ dimension and a synthetic dimension encoded in the atomic spin. Figure adapted from~\citep{mancini2015observation}.}
    \label{fig:syntheticD}
\end{figure}

While chiral edge states have been the major workhorse to characterize topological photonic systems~\citep{ozawaRMP2019topological}, a relatively limited number of experiments have investigated edge currents in atomic systems. 
On the one hand, this is due to the difficulty of directly measuring currents in cold-atom systems, though significant theoretical and experimental advances have been achieved in recent years \citep{Kessler2014,Hauke2014,Krinner2015,Laflamme2017,Scherg2018,Brown2019,Nichols2019,Jepsen2020,Geier2021}. 
On the other hand, an obstacle is the absence of sharp boundaries in standard, harmonically trapped atomic clouds. Also for this reason, the remarkable studies of edge transport in Ref.~\citep{mancini2015observation,stuhl2015visualizing} illustrated in Fig.~\ref{fig:syntheticD} made use of a synthetic dimension set-up combining a spatially one-dimensional lattice with an additional dimension formed by the internal spin degrees. 

A great advantage of atomic systems is the possibility of combining macroscopic transport measurements with a direct microscopic insight into the quantum many-body wavefunction \citep{Gebert2020}. In the IQH context, a remarkable example in this direction was the experimental tomography of the Berry phase~\citep{Hauke2014,flaschner2016experimental}, with follow-ups such as the identification of dynamical topological invariants in the time evolution of an atomic system \citep{Tarnowski2019}. 

\subsection{Extensions}

An important extension of the IQH effect is obtained when a spin degree of freedom of the particles is included, $\sigma=\uparrow,\downarrow$, such that Eq.~\eqref{eq:Hkinlatt} becomes 
$H_\mathrm{kin}=-J \sum_{\braket{i,j}}\sum_\sigma \psi_{i,\sigma}^\dagger U_{i,j}^\sigma \psi_{j,\sigma}$. 
In the so-called spin-Hall effect \citep{hasan2010colloquium}, particles with different spin orientations see opposite magnetic fields, $(U_{i,j}^\uparrow)^\dagger=U_{i,j}^\downarrow$ and time-reversal symmetry is restored. 
The result is a system without any net charge current but chiral spin edge currents that run around the sample in opposite directions, see Fig.~\ref{fig:IQHE-spinHE}(b). 
Pioneering proposals have been based on extending schemes for light-induced vector potentials to several  hyperfine levels \citep{Zhu2006} and on state-dependent, slowly time-varying magnetic potentials engineered in an atom chip on a micron scale~\citep{Goldman2010}. 
The spin Hall effect has been observed in free space, where spin-dependent Lorentz forces have been measured using a spatially inhomogeneous spin-orbit coupling field \citep{Beeler2013}, as well as in an optical lattice using laser-assisted tunneling induced in a tilted optical lattice via periodic driving 
\citep{aidelsburger2013realization}.

This physics can be generalized even further by replacing the spin-dependent Peierls phase by a matrix, $H_\mathrm{kin}=-J \sum_{\braket{i,j}}\sum_{\sigma,\sigma'} \psi_{i,\sigma}^\dagger U_{i,j}^{\sigma,\sigma'} \psi_{j,\sigma'}$, where $U_{i,j}^{\sigma,\sigma'}$ mixes the different spin states during hopping. 
Genuinely non-Abelian effects occur if the product of the $U_{i,j}^{\sigma,\sigma'}$ around a closed path (the non-Abelian Wegner--Wilson loop) does not reduce to a simple phase factor $\mathrm{e}^{i\phi} \mathbb{1}$. 
Such a situation can be obtained through periodically shaken optical superlattices where an internal degree of freedom of the atoms plays the role of the (pseudo-)spin  \citep{Hauke2012}. 
As a striking consequence of the non-Abelian Hall effect, novel fractal features have been predicted to appear in the phase diagram \citep{Goldman2009,Bermudez2010}. 

Another exciting direction of research concerns the extension of IQH effects to higher-dimensional, 4+1 geometries as first proposed in Ref.~\citep{price2015four} using the concept of synthetic dimensions mentioned above. Pioneering experimental evidence of the role of higher-dimensional topological invariants in the atomic transport properties under crossed synthetic-electric and magnetic fields in 4-dimensional models was reported in Ref.~\citep{lohse2018exploring}.

\section{Fractional Quantum Hall effect in cold atoms}
\label{sec:FQH}

In the previous section, we have seen the intriguing effects that appear in the presence of completely occupied Landau levels or topological bands, but where interactions play at most a subdominant role. 
Even more subtle physics occurs when the macroscopic degeneracy of partially filled Landau levels is lifted by strong inter-particle interactions and the lowest energy state acquires a non-trivial topology for the many-body wavefunction~\citep{wen1995topological,Kitaev2006,levin2006detecting,Wen2013}.

One of the most celebrated examples of such topological many-body phases of matter is the so-called fractional quantum Hall (FQH) effect, first detected in two-dimensional electron gases under strong magnetic fields as a precise quantization of the transverse conductivity at rational values proportional to the electron filling, i.e., the number of electrons per unit magnetic flux piercing the system~\citep{Tong:QHbook,yoshioka2002quantum,Girvin}. 
As even more intriguing signatures of the many-body topology of the FQH state, excitations with a fractional charge and fractional statistics have been anticipated for these systems, the so-called {\em anyons}~\citep{stern2008anyons}. While fractionally charged edge excitations have been observed in shot-noise measurements~\citep{depicciotto1998direct,saminadayar1997observation}, the observation of fractional statistical properties interpolating between bosonic and fermionic behaviours is still the subject of intense experimental work~\citep{bartolomei2020fractional,nakamura2020direct}.

\begin{figure}
    \centering
    \includegraphics[width=0.9\columnwidth,trim={0 7.35cm 0 0},clip]{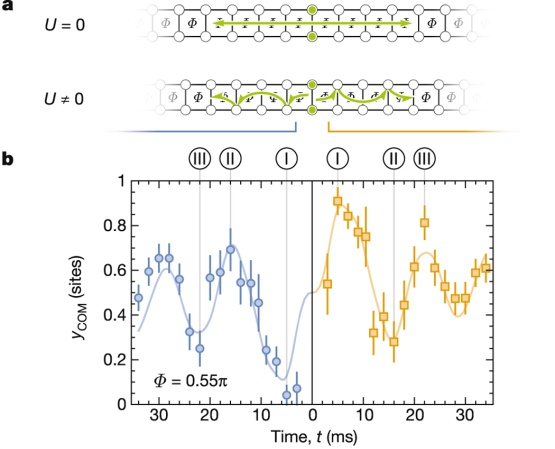}
    \caption{Chiral motion of two interacting atoms in a Harper--Hofstadter ladder, where each plaquette is pierced by a synthetic magnetic flux $\Phi$. Upper panel: sketch of the trajectories for non-interacting ($U=0$) and interacting ($U\neq 0$) atoms. Lower panel: temporal dynamics along $y$ displaying the chiral motion of the so-called skipping orbits. Figure adapted from~\citep{tai2017microscopy}.}
    \label{fig:anyons}
\end{figure}

First steps towards the realization of FQH fluids using ultracold atomic clouds were made by pushing the rotation frequency towards the trapping frequency, so that the atomic cloud expands in space under the effect of the centrifugal potential and reaches the required magnetic flux per particle values. For sufficiently low temperatures, it is then predicted to enter a FQH state~\citep{cooper2008rapidly}. In spite of early concerns about heating due to spurious asymmetries of the trap potential and the need for a fine tuning of the rotation speed, remarkable advances on condensates in rotating traps have been recently reported~\citep{Fletcher2021,Mukherjee2022}. Besides these works on a mean-field-interacting dynamics, hints of correlated FQH physics have been reported in small atomic clusters trapped at the rotating minima of a temporally modulated optical lattice potential~\citep{gemelke2010rotating}.

A different strategy to realize a strongly interacting Harper--Hofstadter lattice using a combination of optical lattice potentials and Raman transitions was adopted in Ref.~\citep{tai2017microscopy} and led to the observation of a correlated two-particle dynamics, a precursor of fractional quantum Hall effects, see Fig.~\ref{fig:anyons}. In the meanwhile, an intense theoretical activity has been devoted to the study of alternative protocols to realize FQH states using, e.g., adiabatic ramps along suitably designed paths in parameter space~\citep{Popp2004,Sorensen2005} or sequences of adiabatic flux insertion and hole replenishing~\citep{Grusdt2014}. 

The main advantage of FQH fluids of ultracold atoms over solid-state systems is the much wider variety of manipulation and diagnostics tools compared to the transport and optical probes traditionally available for electronic systems. While waiting for macroscopic FQH clouds to be experimentally realized, theoretical proposals to investigate their intriguing properties include interferometric techniques~\citep{paredes20011,grusdt2016interferometric}, quasi-hole dynamics in response to external potentials~\citep{Grass2012}, microscopic imaging of the quasi-hole density profile~\citep{macaluso2020charge}, quantum mechanics of anyonic molecules formed by an impurity bound to a FQH quasi-hole~\citep{Zhang2014,lundholm2016emergence,de2020anyonic,yakaboylu2020quantum,Baldelli2021}, the center-of-mass motion of the cloud~\citep{repellin2020fractional}, its circular-dichroic response to circular excitations in the bulk~\citep{repellin2019detecting}, the linear and nonlinear response of the edge modes~\citep{nardin2022linear}. 

Together with direct measurements of the entanglement entropy~\citep{li2008entanglement,jiang2012identifying} as experimentally pioneered in Ref.~\citep{islam2015measuring}, we  anticipate that these proposals will eventually be a most valuable tool to obtain a deeper insight in the basic physics of FQH fluids and synthetic topological matter, and to exploit the benefits of their atomic implementation in view of quantum information tasks~\citep{Nayak:RMP2008}.

\section{Dynamical gauge fields}
\label{sec:LGTs}

In all of the above discussions, the vector potential appearing in Eqs.~\eqref{eq:Hkincont} and~\eqref{eq:Hkinlatt} was assumed to have a constant value, corresponding to a classical, externally imposed background field. 
Entirely new effects can appear once the vector potential is imbued with its own dynamics.

\subsection{Dynamical vector potentials and Peierls phases}

\begin{figure}
	\centering
	\includegraphics[width=0.49\columnwidth]{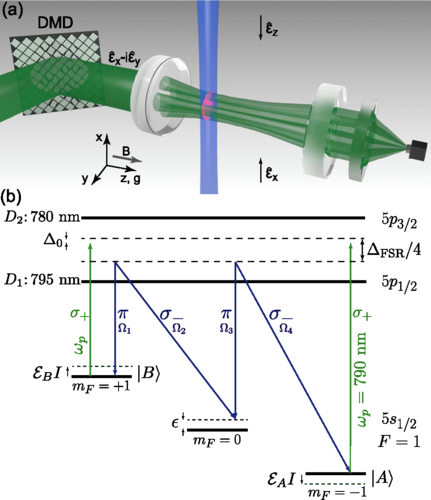}
	\includegraphics[width=0.49\columnwidth]{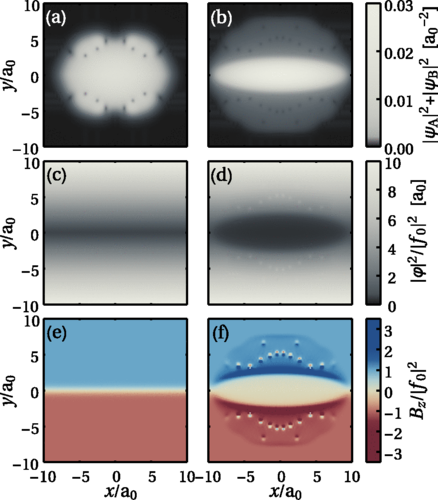}
	\caption{Left: By suitably coupling the internal levels of ultracold atoms to a multi-mode cavity, a backaction from atoms onto the synthetic magnetic field can be engineered, where the spatial profile of the synthetic magnetic field varies depending on the atomic density. Right: Simulations of the atom-cavity system without (panels a,c,e) and with backaction (b,d,f). The backaction leads to an analog of the Meissner effect, where the magnetic field is expelled from the sample, leading to an extended region with $B=0$. Figure adapted from~\citep{Ballantine2017}.}
	\label{fig:dynamicalVectorPotential}
\end{figure}

In the first proposal for continuum systems~\cite{Edmonds2013}, a density-dependent vector potential was anticipated to arise from interaction-induced energy shifts of the internal atomic states involved in the optical transitions generating the Berry phase. Among the observable consequences, density-dependent persistent currents in ring geometries and
chiral solitons were pointed out. Along similar lines, \citep{Ballantine2017} proposed to replace one of the optical fields generating the Berry phase by a multimode cavity field. 
The density of the atoms can then act back on the cavity field, which---due to its multimode character---can acquire a spatial dependence. 
In this way, the equivalent of the celebrated Meissner effect can be generated, whereby a magnetic field is expelled from the atom-cloud sample, see Fig.~\ref{fig:dynamicalVectorPotential}. 

In lattice systems, a complementary direction is taken by turning the Peierls phase into a quantum-mechanical operator. 
This can happen, e.g., through density-dependent tunneling, where the complex hopping phase $U_{i,j}$ becomes a function $U_{i,j}(\hat{n}_i,\hat{n}_j)$ of the atom number operators $\hat{n}_\ell=a_\ell^\dagger a_\ell$ at adjacent sites $\ell=i,j$. 
Such a situation can be generated by Floquet engineering \citep{Meinert2016}, as experimentally demonstrated in Ref.~\citep{Goerg2019}, and it can naturally appear in regimes of very strong interactions or high occupation numbers~\citep{Dutta2011}. Such a density-dependent tunneling has been proposed to lead to deformations of phase diagrams and the generation of novel phases such as pair superfluids \citep{Dutta2011,Maik2013,Johnstone2019,Suthar2020} as well as to topological transitions in the ground state \citep{Raventos2016}.  

In other schemes, the role of the Peierls phase can also be taken over by a second, independent atomic species. E.g., in an atomic mixture, an itinerant species can be coupled to the (pseudo-)spin degree of freedom of a second, kinetically frozen species \citep{Mil2020,Kasper2021}. 
By replacing the static hopping matrix by an additional quantum mechanical spin degree of freedom, $U_{i,j}\rightarrow S_{i,j}^\alpha$, $\alpha=x,y,z$, backaction phenomena between two different fields can lead, e.g., to topological and fractionalization effects such as a Peierls transition where an alternating pattern of strong and weak bonds dynamically emerges similar to polyacetylene \citep{Gonzalez-Cuadra2019,Gonzalez-Cuadra2020}. 
The additional possibilities offered by these experiments with atomic mixtures also enabled the observation of a Lamb shift effect in the atomic self energy.  Similarly to the Lamb shift experienced by QED electrons due to their interaction with the quantum mechanical vector potential, an energy shift is induced here on one atomic species by their interactions with a second one \citep{Rentrop2016}.

\subsection{Lattice gauge theories}

In these approaches, the ``vector potential" acquires its own dynamics, but the combined system of ``particles" and ``vector potential" does not necessarily respect the fundamental laws of QED, making it necessary to take similarities to magnetic fields with a grain of salt. 
To recover a full-fledged system of charged matter moving in a quantum mechanical electromagnetic field, additional care needs to be taken to enforce the $U(1)$ gauge symmetry that interlinks the two different types of dynamical degrees of freedom. 
In particular, the dynamics of the system needs to conserve the local symmetry that underpins QED, given by Gauss' law, at all spatial coordinates and throughout the entire evolution time. 

Although this task is highly challenging, groundbreaking experiments with local gauge symmetry have already been realized in Rydberg quantum simulators \citep{Bernien2017,Surace2020}, using Floquet techniques in optical lattices \citep{Schweizer2019}, exploiting angular momentum conservation in atomic mixtures \citep{Mil2020} (proposed, e.g., in Refs.~\citep{Zohar2013,Stannigel2014,Zache2018}), as well as with energy penalties in optical superlattices \citep{Yang2020,Halimeh2020,Zhou2021}. Pioneering realizations have also been demonstrated in other platforms, in particular trapped ions \citep{Martinez2016,Kokail2019,Nguyen2021}, superconducting qubits \citep{Klco2018,Klco2020,Wang2021,Mildenberger2022}, and even classical electronic circuits \citep{Riechert2021}. 

Most of these implementations focused on salient physical phenomena in one spatial dimension, and it is a current challenge to advance ultracold-atom lattice gauge theories into higher spatial dimensions \citep{Zohar2022}. 
In the future, these may enable, e.g., the investigation of strong-field effects on anomalous currents \citep{Ott2020} as well as a range of new anomalous transport phenomena 
including the chiral magnetic effect \citep{Fukushima2008,Kharzeev2014}, the chiral vortical effect \citep{Son2004}, or the conformal magnetic edge effect \citep{Chu2018,Chernodub2019}.

\section{Conclusions and Perspectives} 
\label{sec:Conclusion}

This review chapter summarized our point of view on the rich physics that can be obtained by synthetic static and dynamical gauge fields in cold-atom setups. 
We have reviewed the way synthetic magnetic fields can be engineered in such systems, discussed progress in controlled implementations of the integer, spin, and fractional quantum Hall effects, as well as some of the fascinating phenomena that appear when the atoms act back on the synthetic magnetic fields, without or with respecting gauge symmetry given by Gauss's law.  
These are, however, just some of the many facets of ongoing research in the broad context of synthetic gauge fields. 

Promising implementations have been designed and realized in other synthetic quantum matter platforms. 
For example, topological photonic systems are presently one of the most active fields of research in optics, with potential technological applications such as efficient optical isolation and large-area, fabrication-disorder-robust topological lasers~\citep{price2022roadmap}. Synthetic magnetic field for photons are one of the key building blocks in this context and, analogously to the atomic case, can be realized in many different ways, from twisted optical cavities leading to an effective overall rotation~\citep{schine2016synthetic}, to static or time-dependent complex hopping phases~\citep{Hafezi:2013NatPhot,Roushan:2016NatPhys}, to Floquet-like temporal modulations~\citep{Rechtsman:2013Nature}. In combination with sizable photon-photon interactions arising from the optical nonlinearity of the underlying optical medium, a synthetic magnetic field for light may lead to the generation of novel states of photonic matter~\citep{carusotto2020photonic}: baby versions of a quantum Hall fluid of light of a few photons have been recently realized~\citep{clark2020observation}.
Other promising platforms are, e.g., analog implementations in trapped ions, where techniques similar to those for optical lattices have been developed to imprint Peierls phases~\citep{Bermudez2011,Manovitz2020}, as well as universal quantum computers, where proposals for realizing fractional quantum Hall states \citep{Rahmani2020} as well as experimental realizations of topological matter exist \citep{nigg2014,Semeghini2021,Satzinger2021}.

All together, the examples discussed in this review illustrate the on-going exciting progress in the study of quantum Hall effects in cold atomic fluids as well as in related synthetic quantum matter platforms~\citep{ozawaprice2019topological}. Beyond the remarkable achievements obtained so far, there are a lot of equally exciting open questions ahead of the community, as well as a number of potentially game-changing applications in quantum technologies \citep{Nayak:RMP2008}.

\bibliography{biblio.bib}

@ARTICLE{Tong:QHbook,
   author = {{Tong}, D.},
    title = {Lectures on the Quantum Hall Effect},
  journal = {ArXiv e-prints},
archivePrefix = "arXiv",
   eprint = {1606.06687},
 primaryClass = "hep-th",
     year = 2016,
    month = jun,
   adsurl = {http://adsabs.harvard.edu/abs/2016arXiv160606687T}
}

@article{Nayak:RMP2008,
  title = {Non-Abelian anyons and topological quantum computation},
  author = {Nayak, Chetan and Simon, Steven H. and Stern, Ady and Freedman, Michael and Das Sarma, Sankar},
  journal = {Rev. Mod. Phys.},
  volume = {80},
  issue = {3},
  pages = {1083--1159},
  numpages = {0},
  year = {2008},
  month = {Sep},
  publisher = {American Physical Society},
  doi = {10.1103/RevModPhys.80.1083},
  url = {https://link.aps.org/doi/10.1103/RevModPhys.80.1083}
}

@article{ozawaRMP2019topological,
  title={Topological photonics},
  author={Ozawa, Tomoki and Price, Hannah M and Amo, Alberto and Goldman, Nathan and Hafezi, Mohammad and Lu, Ling and Rechtsman, Mikael C and Schuster, David and Simon, Jonathan and Zilberberg, Oded and others},
  journal={Reviews of Modern Physics},
  volume={91},
  number={1},
  pages={015006},
  year={2019},
  publisher={APS}
}

@article{ozawaprice2019topological,
  title={Topological quantum matter in synthetic dimensions},
  author={Ozawa, Tomoki and Price, Hannah M},
  journal={Nature Reviews Physics},
  volume={1},
  number={5},
  pages={349--357},
  year={2019},
  publisher={Nature Publishing Group}
}

@article{carusotto2013quantum,
  title={Quantum fluids of light},
  author={Carusotto, Iacopo and Ciuti, Cristiano},
  journal={Reviews of Modern Physics},
  volume={85},
  number={1},
  pages={299},
  year={2013},
  publisher={APS}
}

@article{carusotto2020photonic,
  title={Photonic materials in circuit quantum electrodynamics},
  author={Carusotto, Iacopo and Houck, Andrew A and Koll{\'a}r, Alicia J and Roushan, Pedram and Schuster, David I and Simon, Jonathan},
  journal={Nature Physics},
  volume={16},
  number={3},
  pages={268--279},
  year={2020},
  publisher={Nature Publishing Group}
}

@article{dalibard2011colloquium,
  title={Colloquium: Artificial gauge potentials for neutral atoms},
  author={Dalibard, Jean and Gerbier, Fabrice and Juzeli{\=u}nas, Gediminas and {\"O}hberg, Patrik},
  journal={Reviews of Modern Physics},
  volume={83},
  number={4},
  pages={1523},
  year={2011},
  publisher={APS}
}

@article{goldman2014light,
  title={Light-induced gauge fields for ultracold atoms},
  author={Goldman, Nathan and Juzeli{\=u}nas, G and {\"O}hberg, Patrik and Spielman, Ian B},
  journal={Reports on Progress in Physics},
  volume={77},
  number={12},
  pages={126401},
  year={2014},
  publisher={IOP Publishing}
}

@article{lin2009synthetic,
  title={Synthetic magnetic fields for ultracold neutral atoms},
  author={Lin, Y-J and Compton, Rob L and Jim{\'e}nez-Garc{\'\i}a, Karina and Porto, James V and Spielman, Ian B},
  journal={Nature},
  volume={462},
  number={7273},
  pages={628--632},
  year={2009},
  publisher={Nature Publishing Group}
}

@article{dum1996gauge,
  title={Gauge structures in atom-laser interaction: Bloch oscillations in a dark lattice},
  author={Dum, R and Olshanii, M},
  journal={Physical review letters},
  volume={76},
  number={11},
  pages={1788},
  year={1996},
  publisher={APS}
}

@article{jaksch2003creation,
  title={Creation of effective magnetic fields in optical lattices: the Hofstadter butterfly for cold neutral atoms},
  author={Jaksch, Dieter and Zoller, Peter},
  journal={New Journal of Physics},
  volume={5},
  number={1},
  pages={56},
  year={2003},
  publisher={IOP Publishing}
}

@article{HaukeEckardt,
  title = {Tunable Gauge Potential for Neutral and Spinless Particles in Driven Optical Lattices},
  author = {Struck, J. and \"Olschl\"ager, C. and Weinberg, M. and Hauke, P. and Simonet, J. and Eckardt, A. and Lewenstein, M. and Sengstock, K. and Windpassinger, P.},
  journal = {Phys. Rev. Lett.},
  volume = {108},
  issue = {22},
  pages = {225304},
  numpages = {5},
  year = {2012},
  month = {May},
  publisher = {American Physical Society},
  doi = {10.1103/PhysRevLett.108.225304},
  url = {https://link.aps.org/doi/10.1103/PhysRevLett.108.225304}
}

@article{jotzu2014experimental,
  title={Experimental realization of the topological Haldane model with ultracold fermions},
  author={Jotzu, Gregor and Messer, Michael and Desbuquois, R{\'e}mi and Lebrat, Martin and Uehlinger, Thomas and Greif, Daniel and Esslinger, Tilman},
  journal={Nature},
  volume={515},
  number={7526},
  pages={237--240},
  year={2014},
  publisher={Nature Publishing Group}
}

@article{flaschner2016experimental,
  title={Experimental reconstruction of the Berry curvature in a Floquet Bloch band},
  author={Fl{\"a}schner, Nick and Rem, BS and Tarnowski, M and Vogel, D and L{\"u}hmann, D-S and Sengstock, K and Weitenberg, Christof},
  journal={Science},
  volume={352},
  number={6289},
  pages={1091--1094},
  year={2016},
  publisher={American Association for the Advancement of Science}
}

@article{celi2014synthetic,
  title={Synthetic gauge fields in synthetic dimensions},
  author={Celi, Alessio and Massignan, Pietro and Ruseckas, Julius and Goldman, Nathan and Spielman, Ian B and Juzeli{\=u}nas, G and Lewenstein, M},
  journal={Physical review letters},
  volume={112},
  number={4},
  pages={043001},
  year={2014},
  publisher={APS}
}

@article{price2015four,
  title={Four-dimensional quantum Hall effect with ultracold atoms},
  author={Price, Hannah M and Zilberberg, Oded and Ozawa, Tomoki and Carusotto, Iacopo and Goldman, Nathan},
  journal={Physical review letters},
  volume={115},
  number={19},
  pages={195303},
  year={2015},
  publisher={APS}
}

@article{mancini2015observation,
  title={Observation of chiral edge states with neutral fermions in synthetic Hall ribbons},
  author={Mancini, Marco and Pagano, Guido and Cappellini, Giacomo and Livi, Lorenzo and Rider, Marie and Catani, Jacopo and Sias, Carlo and Zoller, Peter and Inguscio, Massimo and Dalmonte, Marcello and others},
  journal={Science},
  volume={349},
  number={6255},
  pages={1510--1513},
  year={2015},
  publisher={American Association for the Advancement of Science}
}

@article{stuhl2015visualizing,
  title={Visualizing edge states with an atomic Bose gas in the quantum Hall regime},
  author={Stuhl, BK and Lu, H-I and Aycock, LM and Genkina, D and Spielman, IB},
  journal={Science},
  volume={349},
  number={6255},
  pages={1514--1518},
  year={2015},
  publisher={American Association for the Advancement of Science}
}

@article{thouless1982quantized,
  title={Quantized Hall conductance in a two-dimensional periodic potential},
  author={Thouless, David J and Kohmoto, Mahito and Nightingale, M Peter and den Nijs, Marcel},
  journal={Physical review letters},
  volume={49},
  number={6},
  pages={405},
  year={1982},
  publisher={APS}
}

@article{hasan2010colloquium,
  title={Colloquium: topological insulators},
  author={Hasan, M Zahid and Kane, Charles L},
  journal={Reviews of modern physics},
  volume={82},
  number={4},
  pages={3045},
  year={2010},
  publisher={APS}
}

@article{dudarev2004spin,
  title={Spin-orbit coupling and berry phase with ultracold atoms in 2d optical lattices},
  author={Dudarev, Artem M and Diener, Roberto B and Carusotto, Iacopo and Niu, Qian},
  journal={Physical review letters},
  volume={92},
  number={15},
  pages={153005},
  year={2004},
  publisher={APS}
}

@article{aidelsburger2015measuring,
  title={Measuring the Chern number of Hofstadter bands with ultracold bosonic atoms},
  author={Aidelsburger, Monika and Lohse, Michael and Schweizer, Christian and Atala, Marcos and Barreiro, Julio T and Nascimb{\`e}ne, Sylvain and Cooper, NR and Bloch, Immanuel and Goldman, Nathan},
  journal={Nature Physics},
  volume={11},
  number={2},
  pages={162--166},
  year={2015},
  publisher={Nature Publishing Group}
}

@article{lohse2018exploring,
  title={Exploring 4D quantum Hall physics with a 2D topological charge pump},
  author={Lohse, Michael and Schweizer, Christian and Price, Hannah M and Zilberberg, Oded and Bloch, Immanuel},
  journal={Nature},
  volume={553},
  number={7686},
  pages={55--58},
  year={2018},
  publisher={Nature Publishing Group}
}

@article{stern2008anyons,
  title={Anyons and the quantum Hall effect—a pedagogical review},
  author={Stern, Ady},
  journal={Annals of Physics},
  volume={323},
  number={1},
  pages={204--249},
  year={2008},
  publisher={Elsevier}
}

@article{depicciotto1998direct,
  title={Direct observation of a fractional charge},
  author={De-Picciotto, R and Reznikov, M and Heiblum, M and Umansky, V and Bunin, G and Mahalu, D},
  journal={Physica B: Condensed Matter},
  volume={249},
  pages={395--400},
  year={1998},
  publisher={Elsevier}
}

@article{bartolomei2020fractional,
  title={Fractional statistics in anyon collisions},
  author={Bartolomei, Hugo and Kumar, Manohar and Bisognin, R{\'e}mi and Marguerite, Arthur and Berroir, J-M and Bocquillon, Erwann and Placais, Bernard and Cavanna, Antonella and Dong, Q and Gennser, Ulf and others},
  journal={Science},
  volume={368},
  number={6487},
  pages={173--177},
  year={2020},
  publisher={American Association for the Advancement of Science}
}

@article{nakamura2020direct,
  title={Direct observation of anyonic braiding statistics},
  author={Nakamura, James and Liang, Shuang and Gardner, Geoffrey C and Manfra, Michael J},
  journal={Nature Physics},
  volume={16},
  number={9},
  pages={931--936},
  year={2020},
  publisher={Nature Publishing Group}
}

@article{tai2017microscopy,
  title={Microscopy of the interacting Harper--Hofstadter model in the two-body limit},
  author={Tai, M Eric and Lukin, Alexander and Rispoli, Matthew and Schittko, Robert and Menke, Tim and Borgnia, Dan and Preiss, Philipp M and Grusdt, Fabian and Kaufman, Adam M and Greiner, Markus},
  journal={Nature},
  volume={546},
  number={7659},
  pages={519--523},
  year={2017},
  publisher={Nature Publishing Group}
}

@article{cooper2008rapidly,
  title={Rapidly rotating atomic gases},
  author={Cooper, Nigel R},
  journal={Advances in Physics},
  volume={57},
  number={6},
  pages={539--616},
  year={2008},
  publisher={Taylor \& Francis}
}

@article{bretin2004fast,
  title={Fast rotation of a Bose-Einstein condensate},
  author={Bretin, Vincent and Stock, Sabine and Seurin, Yannick and Dalibard, Jean},
  journal={Physical review letters},
  volume={92},
  number={5},
  pages={050403},
  year={2004},
  publisher={APS}
}

@article{gemelke2010rotating,
  title={Rotating few-body atomic systems in the fractional quantum Hall regime},
  author={Gemelke, Nathan and Sarajlic, Edina and Chu, Steven},
  journal={arXiv preprint arXiv:1007.2677},
  year={2010}
}

@article{paredes20011,
  title={1 2-anyons in small atomic Bose-Einstein condensates},
  author={Paredes, B and Fedichev, P and Cirac, JI and Zoller, P},
  journal={Physical review letters},
  volume={87},
  number={1},
  pages={010402},
  year={2001},
  publisher={APS}
}

@article{grusdt2016interferometric,
  title={Interferometric measurements of many-body topological invariants using mobile impurities},
  author={Grusdt, Fabian and Yao, Norman Y and Abanin, D and Fleischhauer, Michael and Demler, E},
  journal={Nature communications},
  volume={7},
  number={1},
  pages={1--9},
  year={2016},
  publisher={Nature Publishing Group}
}

@article{de2020anyonic,
  title={Anyonic molecules in atomic fractional quantum Hall liquids: a quantitative probe of fractional charge and anyonic statistics},
  author={de las Heras, A Mu{\~n}oz and Macaluso, Elia and Carusotto, Iacopo},
  journal={Physical Review X},
  volume={10},
  number={4},
  pages={041058},
  year={2020},
  publisher={APS}
}

@article{macaluso2020charge,
  title={Charge and statistics of lattice quasiholes from density measurements: A tree tensor network study},
  author={Macaluso, Elia and Comparin, Tommaso and Umucal{\i}lar, Rifat Onur and Gerster, Matthias and Montangero, Simone and Rizzi, Matteo and Carusotto, Iacopo},
  journal={Physical review research},
  volume={2},
  number={1},
  pages={013145},
  year={2020},
  publisher={APS}
}

@article{lundholm2016emergence,
  title={Emergence of fractional statistics for tracer particles in a Laughlin liquid},
  author={Lundholm, Douglas and Rougerie, Nicolas},
  journal={Physical Review Letters},
  volume={116},
  number={17},
  pages={170401},
  year={2016},
  publisher={APS}
}

@article{yakaboylu2020quantum,
  title={Quantum impurity model for anyons},
  author={Yakaboylu, Enderalp and Ghazaryan, Areg and Lundholm, Douglas and Rougerie, Nicolas and Lemeshko, Mikhail and Seiringer, Robert},
  journal={Physical Review B},
  volume={102},
  number={14},
  pages={144109},
  year={2020},
  publisher={APS}
}

@article{repellin2020fractional,
  title={Fractional Chern insulators of few bosons in a box: Hall plateaus from center-of-mass drifts and density profiles},
  author={Repellin, Cecile and Leonard, Julian and Goldman, Nathan},
  journal={Physical Review A},
  volume={102},
  number={6},
  pages={063316},
  year={2020},
  publisher={APS}
}

@article{repellin2019detecting,
  title={Detecting fractional Chern insulators through circular dichroism},
  author={Repellin, Cecile and Goldman, Nathan},
  journal={Physical review letters},
  volume={122},
  number={16},
  pages={166801},
  year={2019},
  publisher={APS}
}

@article{price2022roadmap,
  title={Roadmap on topological photonics},
  author={Price, Hannah and Chong, Yidong and Khanikaev, Alexander and Schomerus, Henning and Maczewsky, Lukas J and Kremer, Mark and Heinrich, Matthias and Szameit, Alexander and Zilberberg, Oded and Yang, Yihao and others},
  journal={Journal of Physics: Photonics},
  year={2022},
  publisher={IOP Publishing}
}

@article{clark2020observation,
  title={Observation of Laughlin states made of light},
  author={Clark, Logan W and Schine, Nathan and Baum, Claire and Jia, Ningyuan and Simon, Jonathan},
  journal={Nature},
  volume={582},
  number={7810},
  pages={41--45},
  year={2020},
  publisher={Nature Publishing Group}
}

@article{schine2016synthetic,
  title={Synthetic Landau levels for photons},
  author={Schine, Nathan and Ryou, Albert and Gromov, Andrey and Sommer, Ariel and Simon, Jonathan},
  journal={Nature},
  volume={534},
  number={7609},
  pages={671--675},
  year={2016},
  publisher={Nature Publishing Group}
}

@article{Roushan:2016NatPhys,
  title={Chiral ground-state currents of interacting photons in a synthetic magnetic field},
  author={Roushan, Pedram and Neill, Charles and Megrant, Anthony and Chen, Yu and Babbush, Ryan and Barends, Rami and Campbell, Brooks and Chen, Zijun and Chiaro, Ben and Dunsworth, Andrew and others},
  journal={Nature Physics},
  volume={13},
  number={2},
  pages={146},
  year={2017},
  url={https://www.nature.com/articles/nphys3930},
  publisher={Nature Publishing Group}
}

@article{Rechtsman:2013Nature,
  title={Photonic Floquet topological insulators},
  author={Rechtsman, Mikael C and Zeuner, Julia M and Plotnik, Yonatan and Lumer, Yaakov and Podolsky, Daniel and Dreisow, Felix and Nolte, Stefan and Segev, Mordechai and Szameit, Alexander},
  journal={Nature},
  volume={496},
  number={7444},
  pages={196--200},
  year={2013},
  url={http://www.nature.com/nature/journal/v496/n7444/full/nature12066.html},
  publisher={Nature Publishing Group}
}

@article{Hafezi:2013NatPhot,
  title={Imaging topological edge states in silicon photonics},
  author={Hafezi, Mohammad and Mittal, S and Fan, J and Migdall, A and Taylor, JM},
  journal={Nature Photonics},
  volume={7},
  number={12},
  pages={1001--1005},
  year={2013},
  url={http://www.nature.com/nphoton/journal/v7/n12/full/nphoton.2013.274.html},
  publisher={Nature Publishing Group}
}

@book{yoshioka2002quantum,
  title={The Quantum Hall Effect},
  author={Yoshioka, D.},
  isbn={9783540431152},
  lccn={02021123},
  series={Physics and astronomy online library},
  url={https://books.google.it/books?id=hFpOcAgCviQC},
  year={2002},
  publisher={Springer}
}

@book{pitaevskii2016bose,
  title={Bose-Einstein Condensation and Superfluidity},
  author={Pitaevski{\u\i}, L.P. and Stringari, S.},
  isbn={9780198758884},
  lccn={2015947456},
  series={International series of monographs on physics},
  url={https://books.google.it/books?id=k\_ZGCwAAQBAJ},
  year={2016},
  publisher={Oxford University Press}
}

@InProceedings{Girvin,
author="Girvin, S. M.",
editor="Comtet, A.
and Jolic{\oe}ur, T.
and Ouvry, S.
and David, F.",
title="The Quantum Hall Effect: Novel Excitations And Broken Symmetries",
booktitle="Aspects topologiques de la physique en basse dimension. Topological aspects of low dimensional systems",
year="1999",
publisher="Springer Berlin Heidelberg",
address="Berlin, Heidelberg",
pages="53--175",
isbn="978-3-540-46637-6"
}

@article{goldman2014periodically,
  title={Periodically driven quantum systems: effective Hamiltonians and engineered gauge fields},
  author={Goldman, Nathan and Dalibard, Jean},
  journal={Physical review X},
  volume={4},
  number={3},
  pages={031027},
  year={2014},
  publisher={APS}
}

@article{miyake2013realizing,
  title={Realizing the Harper Hamiltonian with laser-assisted tunneling in optical lattices},
  author={Miyake, Hirokazu and Siviloglou, Georgios A and Kennedy, Colin J and Burton, William Cody and Ketterle, Wolfgang},
  journal={Physical review letters},
  volume={111},
  number={18},
  pages={185302},
  year={2013},
  publisher={APS}
}

@article{aidelsburger2013realization,
  title={Realization of the Hofstadter Hamiltonian with ultracold atoms in optical lattices},
  author={Aidelsburger, Monika and Atala, Marcos and Lohse, Michael and Barreiro, Julio T and Paredes, B and Bloch, Immanuel},
  journal={Physical review letters},
  volume={111},
  number={18},
  pages={185301},
  year={2013},
  publisher={APS}
}

@article{abo2001observation,
  title={Observation of vortex lattices in Bose-Einstein condensates},
  author={Abo-Shaeer, Jamil R and Raman, Chandra and Vogels, Johnny M and Ketterle, Wolfgang},
  journal={Science},
  volume={292},
  number={5516},
  pages={476--479},
  year={2001},
  publisher={American Association for the Advancement of Science}
}

@article{schweikhard2004rapidly,
  title={Rapidly rotating Bose-Einstein condensates in and near the lowest Landau level},
  author={Schweikhard, Volker and Coddington, Ian and Engels, Peter and Mogendorff, Veronique P and Cornell, Eric A},
  journal={Physical review letters},
  volume={92},
  number={4},
  pages={040404},
  year={2004},
  publisher={APS}
}

@article{nardin2022linear,
  title={Linear and nonlinear edge dynamics of trapped fractional quantum Hall droplets beyond the chiral Luttinger liquid paradigm},
  author={Nardin, Alberto and Carusotto, Iacopo},
  journal={arXiv preprint arXiv:2203.02539},
  year={2022}
}

@article{wen1995topological,
  title={Topological orders and edge excitations in fractional quantum Hall states},
  author={Wen, Xiao-Gang},
  journal={Advances in Physics},
  volume={44},
  number={5},
  pages={405--473},
  year={1995},
  publisher={Taylor \& Francis}
}

@article{levin2006detecting,
  title={Detecting topological order in a ground state wave function},
  author={Levin, Michael and Wen, Xiao-Gang},
  journal={Physical review letters},
  volume={96},
  number={11},
  pages={110405},
  year={2006},
  publisher={APS}
}

@article{saminadayar1997observation,
  title={Observation of the e/3 fractionally charged Laughlin quasiparticle},
  author={Saminadayar, L and Glattli, DC and Jin, Y and Etienne, B c-m},
  journal={Physical Review Letters},
  volume={79},
  number={13},
  pages={2526},
  year={1997},
  publisher={APS}
}

@article{islam2015measuring,
  title={Measuring entanglement entropy in a quantum many-body system},
  author={Islam, Rajibul and Ma, Ruichao and Preiss, Philipp M and Eric Tai, M and Lukin, Alexander and Rispoli, Matthew and Greiner, Markus},
  journal={Nature},
  volume={528},
  number={7580},
  pages={77--83},
  year={2015},
  publisher={Nature Publishing Group}
}

@article{li2008entanglement,
  title={Entanglement spectrum as a generalization of entanglement entropy: Identification of topological order in non-abelian fractional quantum hall effect states},
  author={Li, Hui and Haldane, F Duncan M},
  journal={Physical review letters},
  volume={101},
  number={1},
  pages={010504},
  year={2008},
  publisher={APS}
}

@article{jiang2012identifying,
  title={Identifying topological order by entanglement entropy},
  author={Jiang, Hong-Chen and Wang, Zhenghan and Balents, Leon},
  journal={Nature Physics},
  volume={8},
  number={12},
  pages={902--905},
  year={2012},
  publisher={Nature Publishing Group}
}

@article{dalibard2015introduction,
  title={Introduction to the physics of artificial gauge fields},
  author={Dalibard, Jean},
  journal={Quantum Matter at Ultralow Temperatures},
  year={2015}
}

@article{LeBlanc_2015,
	doi = {10.1088/1367-2630/17/6/065016},
	url = {https://doi.org/10.1088/1367-2630/17/6/065016},
	year = 2015,
	month = {jun},
	publisher = {{IOP} Publishing},
	volume = {17},
	number = {6},
	pages = {065016},
	author = {L J LeBlanc and K Jim{\'{e}}nez-Garc{\'{\i}}a and R A Williams and M C Beeler and W D Phillips and I B Spielman},
	title = {Gauge matters: observing the vortex-nucleation transition in a Bose condensate},
	journal = {New Journal of Physics}
}

@article{nigg2014,
  title={Quantum computations on a topologically encoded qubit},
  author={D. Nigg and M. Mueller and E. A. Martinez and P. Schindler and M. Hennrich and T. Monz and M. A. Martin-Delgado and R. Blatt},
  journal={Science},
  year={2014}, 
  volume = {345},
	pages = {302-305},
}

@article{Manovitz2020,
  title = {Quantum Simulations with Complex Geometries and Synthetic Gauge Fields in a Trapped Ion Chain},
  author = {Manovitz, Tom and Shapira, Yotam and Akerman, Nitzan and Stern, Ady and Ozeri, Roee},
  journal = {PRX Quantum},
  volume = {1},
  issue = {2},
  pages = {020303},
  numpages = {13},
  year = {2020},
  month = {Oct},
  publisher = {American Physical Society},
  doi = {10.1103/PRXQuantum.1.020303},
  url = {https://link.aps.org/doi/10.1103/PRXQuantum.1.020303}
}

@article{Bermudez2011,
  title = {Synthetic Gauge Fields for Vibrational Excitations of Trapped Ions},
  author = {Bermudez, Alejandro and Schaetz, Tobias and Porras, Diego},
  journal = {Phys. Rev. Lett.},
  volume = {107},
  issue = {15},
  pages = {150501},
  numpages = {5},
  year = {2011},
  month = {Oct},
  publisher = {American Physical Society},
  doi = {10.1103/PhysRevLett.107.150501},
  url = {https://link.aps.org/doi/10.1103/PhysRevLett.107.150501}
}

@article{Kitaev2006,
  title = {Topological Entanglement Entropy},
  author = {Kitaev, Alexei and Preskill, John},
  journal = {Phys. Rev. Lett.},
  volume = {96},
  issue = {11},
  pages = {110404},
  numpages = {4},
  year = {2006},
  month = {Mar},
  publisher = {American Physical Society},
  doi = {10.1103/PhysRevLett.96.110404},
  url = {https://link.aps.org/doi/10.1103/PhysRevLett.96.110404}
}

@article{Wen2013,
  title = {Topological Order: From Long-Range Entangled Quantum Matter to a Unified Origin of Light and Electrons},
  author = {X.-G. Wen},
  journal = {International Scholarly Research Notices},
  year = {2013},
  pages = {198710}
}

@article{Hauke2012,
  title = {Non-Abelian Gauge Fields and Topological Insulators in Shaken Optical Lattices},
  author = {Hauke, Philipp and Tieleman, Olivier and Celi, Alessio and \"Olschl\"ager, Christoph and Simonet, Juliette and Struck, Julian and Weinberg, Malte and Windpassinger, Patrick and Sengstock, Klaus and Lewenstein, Maciej and Eckardt, Andr\'e},
  journal = {Phys. Rev. Lett.},
  volume = {109},
  issue = {14},
  pages = {145301},
  numpages = {6},
  year = {2012},
  month = {Oct},
  publisher = {American Physical Society},
  doi = {10.1103/PhysRevLett.109.145301},
  url = {https://link.aps.org/doi/10.1103/PhysRevLett.109.145301}
}

@article{Oka2009,
  title = {Photovoltaic Hall effect in graphene},
  author = {Oka, Takashi and Aoki, Hideo},
  journal = {Phys. Rev. B},
  volume = {79},
  issue = {8},
  pages = {081406},
  numpages = {4},
  year = {2009},
  month = {Feb},
  publisher = {American Physical Society},
  doi = {10.1103/PhysRevB.79.081406},
  url = {https://link.aps.org/doi/10.1103/PhysRevB.79.081406}
}

@Article{Bukov2015,
  author    = {Marin Bukov and Luca D'Alessio and Anatoli Polkovnikov},
  journal   = {Advances in Physics},
  title     = {Universal high-frequency behavior of periodically driven systems: from dynamical stabilization to Floquet engineering},
  year      = {2015},
  number    = {2},
  pages     = {139-226},
  volume    = {64},
  doi       = {10.1080/00018732.2015.1055918},
  eprint    = {https://doi.org/10.1080/00018732.2015.1055918},
  publisher = {Taylor & Francis},
  url       = {https://doi.org/10.1080/00018732.2015.1055918},
}

@article{Kolovsky2011,
	Author = {A. R. Kolovsky},
	Journal = {Euro. Phys. Lett.},
	title = {Creating artificial magnetic fields for cold atoms by photon-assisted tunneling}, 
	Pages = {20003},
	Volume = {93},
	Year = {2011}
}

@article{Kitagawa2011,
  title = {Transport properties of nonequilibrium systems under the application of light: Photoinduced quantum Hall insulators without Landau levels},
  author = {Kitagawa, Takuya and Oka, Takashi and Brataas, Arne and Fu, Liang and Demler, Eugene},
  journal = {Phys. Rev. B},
  volume = {84},
  issue = {23},
  pages = {235108},
  numpages = {13},
  year = {2011},
  month = {Dec},
  publisher = {American Physical Society},
  doi = {10.1103/PhysRevB.84.235108},
  url = {https://link.aps.org/doi/10.1103/PhysRevB.84.235108}
}

@article{Kitagawa2010,
  title = {Topological characterization of periodically driven quantum systems},
  author = {Kitagawa, Takuya and Berg, Erez and Rudner, Mark and Demler, Eugene},
  journal = {Phys. Rev. B},
  volume = {82},
  issue = {23},
  pages = {235114},
  numpages = {12},
  year = {2010},
  month = {Dec},
  publisher = {American Physical Society},
  doi = {10.1103/PhysRevB.82.235114},
  url = {https://link.aps.org/doi/10.1103/PhysRevB.82.235114}
}

@article{Eckardt2015,
  title = {High-frequency approximation for periodically driven quantum systems from a Floquet-space perspective},
  author = {Andre Eckardt and Egidijus Anisimovas},
  journal = {New J. Phys.},
  volume = {17},
  pages = {093039},
  year = {2015}
}

@article{Geier2021Exciting,
  title = {Exciting the Goldstone Modes of a Supersolid Spin-Orbit-Coupled Bose Gas},
  author = {Geier, Kevin T. and Martone, Giovanni I. and Hauke, Philipp and Stringari, Sandro},
  journal = {Phys. Rev. Lett.},
  volume = {127},
  issue = {11},
  pages = {115301},
  numpages = {6},
  year = {2021},
  month = {Sep},
  publisher = {American Physical Society},
  doi = {10.1103/PhysRevLett.127.115301},
  url = {https://link.aps.org/doi/10.1103/PhysRevLett.127.115301}
}

@article{Li2017,
  title = {A stripe
phase with supersolid properties in spin–orbit-coupled
Bose–Einstein condensates},
  author = {J.-R. Li and J. Lee and W. Huang and S. Burchesky and B. Shteynas and F. C. Top and A. O. Jamison and W. Ketterle},
  journal = {Nature (London)},
  volume = {543},
  pages = {91},
  year = {2017},
}

@article{Putra2020,
  title = {Spatial Coherence of Spin-Orbit-Coupled Bose Gases},
  author = {Putra, Andika and Salces-C\'arcoba, F. and Yue, Yuchen and Sugawa, Seiji and Spielman, I. B.},
  journal = {Phys. Rev. Lett.},
  volume = {124},
  issue = {5},
  pages = {053605},
  numpages = {6},
  year = {2020},
  month = {Feb},
  publisher = {American Physical Society},
  doi = {10.1103/PhysRevLett.124.053605},
  url = {https://link.aps.org/doi/10.1103/PhysRevLett.124.053605}
}

@article{Hauke2014,
  title = {Tomography of Band Insulators from Quench Dynamics},
  author = {Hauke, Philipp and Lewenstein, Maciej and Eckardt, Andr\'e},
  journal = {Phys. Rev. Lett.},
  volume = {113},
  issue = {4},
  pages = {045303},
  numpages = {5},
  year = {2014},
  month = {Jul},
  publisher = {American Physical Society},
  doi = {10.1103/PhysRevLett.113.045303},
  url = {https://link.aps.org/doi/10.1103/PhysRevLett.113.045303}
}

@article{Atala2012,
	Author = {Marcos Atala and Monika Aidelsburger and Julio T. Barreiro and Dmitry Abanin and Takuya Kitagawa and Eugene Demler and Immanuel Bloch},
	Journal = {Nat. Phys.},
	Pages = {795},
	Title = {Direct Measurement of the {Z}ak phase in Topological Bloch Bands},
	Volume = {9},
	Year = {2013}
}

@article{Gebert2020,
  title = {Local Chern marker of smoothly confined Hofstadter fermions},
  author = {Gebert, Urs and Irsigler, Bernhard and Hofstetter, Walter},
  journal = {Phys. Rev. A},
  volume = {101},
  issue = {6},
  pages = {063606},
  numpages = {7},
  year = {2020},
  month = {Jun},
  publisher = {American Physical Society},
  doi = {10.1103/PhysRevA.101.063606},
  url = {https://link.aps.org/doi/10.1103/PhysRevA.101.063606}
}

@article{Goldman2009,
  title = {Ultracold atomic gases in non-Abelian gauge potentials: The case of constant Wilson loop},
  author = {Goldman, N. and Kubasiak, A. and Gaspard, P. and Lewenstein, M.},
  journal = {Phys. Rev. A},
  volume = {79},
  issue = {2},
  pages = {023624},
  numpages = {11},
  year = {2009},
  month = {Feb},
  publisher = {American Physical Society},
  doi = {10.1103/PhysRevA.79.023624},
  url = {https://link.aps.org/doi/10.1103/PhysRevA.79.023624}
}

@article{Bermudez2010,
	Author = {Bermudez, A. and Goldman, N. and Kubasiak, A. and Lewenstein, M. and Martin-Delgado, M.},
	Journal = {New J. Phys.},
	Pages = {033041},
	Title = {Topological phase transitions in the non-{A}belian honeycomb lattice},
	Volume = {12},
	Year = {2010}
}

@article{Goldman2010,
  title = {Realistic Time-Reversal Invariant Topological Insulators with Neutral Atoms},
  author = {Goldman, N. and Satija, I. and Nikolic, P. and Bermudez, A. and Martin-Delgado, M. A. and Lewenstein, M. and Spielman, I. B.},
  journal = {Phys. Rev. Lett.},
  volume = {105},
  issue = {25},
  pages = {255302},
  numpages = {4},
  year = {2010},
  month = {Dec},
  publisher = {American Physical Society},
  doi = {10.1103/PhysRevLett.105.255302},
  url = {https://link.aps.org/doi/10.1103/PhysRevLett.105.255302}
}

@article{Zhu2006,
  title = {Spin Hall Effects for Cold Atoms in a Light-Induced Gauge Potential},
  author = {Zhu, Shi-Liang and Fu, Hao and Wu, C.-J. and Zhang, S.-C. and Duan, L.-M.},
  journal = {Phys. Rev. Lett.},
  volume = {97},
  issue = {24},
  pages = {240401},
  numpages = {4},
  year = {2006},
  month = {Dec},
  publisher = {American Physical Society},
  doi = {10.1103/PhysRevLett.97.240401},
  url = {https://link.aps.org/doi/10.1103/PhysRevLett.97.240401}
}

@article{Beeler2013,
  title = {The spin Hall effect in a quantum gas},
  author = {M. C. Beeler and R. A. Williams and K. Jiménez-García and L. J. LeBlanc and A. R. Perry and I. B. Spielman},
  journal = {Nature},
  volume = {498},
  pages = {201--204},
  numpages = {4},
  year = {2013}
}

@article{Ballantine2017,
  title = {Meissner-like Effect for a Synthetic Gauge Field in Multimode Cavity QED},
  author = {Ballantine, Kyle E. and Lev, Benjamin L. and Keeling, Jonathan},
  journal = {Phys. Rev. Lett.},
  volume = {118},
  issue = {4},
  pages = {045302},
  numpages = {7},
  year = {2017},
  month = {Jan},
  publisher = {American Physical Society},
  doi = {10.1103/PhysRevLett.118.045302},
  url = {https://link.aps.org/doi/10.1103/PhysRevLett.118.045302}
}

@article{Gonzalez-Cuadra2020,
  title = {Dynamical Solitons and Boson Fractionalization in Cold-Atom Topological Insulators},
  author = {Gonz\'alez-Cuadra, D. and Dauphin, A. and Grzybowski, P. R. and Lewenstein, M. and Bermudez, A.},
  journal = {Phys. Rev. Lett.},
  volume = {125},
  issue = {26},
  pages = {265301},
  numpages = {6},
  year = {2020},
  month = {Dec},
  publisher = {American Physical Society},
  doi = {10.1103/PhysRevLett.125.265301},
  url = {https://link.aps.org/doi/10.1103/PhysRevLett.125.265301}
}

@article{Gonzalez-Cuadra2019,
  title = {Intertwined Topological Phases induced by Emergent Symmetry Protection},
  author = {Gonz\'alez-Cuadra, D. and Bermudez, A. and Grzybowski, P. R. and Lewenstein, M. and Dauphin, A.},
  journal = {Nat. Commun.},
  volume = {10},
  pages = {2694},
  year = {2019},
  doi = {https://doi.org/10.1038/s41467-019-10796-8},
  url = {https://doi.org/10.1038/s41467-019-10796-8}
}

@article{Rentrop2016,
  title = {Observation of the Phononic Lamb Shift with a Synthetic Vacuum},
  author = {Rentrop, T. and Trautmann, A. and Olivares, F. A. and Jendrzejewski, F. and Komnik, A. and Oberthaler, M. K.},
  journal = {Phys. Rev. X},
  volume = {6},
  issue = {4},
  pages = {041041},
  numpages = {10},
  year = {2016},
  month = {Nov},
  publisher = {American Physical Society},
  doi = {10.1103/PhysRevX.6.041041},
  url = {https://link.aps.org/doi/10.1103/PhysRevX.6.041041}
}

@article{Bernien2017,
	Author = {Bernien, Hannes and Schwartz, Sylvain and Keesling, Alexander and Levine, Harry and Omran, Ahmed and Pichler, Hannes and Choi, Soonwon and Zibrov, Alexander S. and Endres, Manuel and Greiner, Markus and Vuleti{\'c}, Vladan and Lukin, Mikhail D.},
	Doi = {10.1038/nature24622},
	Id = {Bernien2017},
	Isbn = {1476-4687},
	Journal = {Nature},
	Number = {7682},
	Pages = {579--584},
	Title = {Probing many-body dynamics on a 51-atom quantum simulator},
	Ty = {JOUR},
	Url = {https://doi.org/10.1038/nature24622},
	Volume = {551},
	Year = {2017}
}

@article{Surace2020,
  title = {Lattice Gauge Theories and String Dynamics in Rydberg Atom Quantum Simulators},
  author = {Surace, Federica M. and Mazza, Paolo P. and Giudici, Giuliano and Lerose, Alessio and Gambassi, Andrea and Dalmonte, Marcello},
  journal = {Phys. Rev. X},
  volume = {10},
  issue = {2},
  pages = {021041},
  numpages = {14},
  year = {2020},
  month = {May},
  publisher = {American Physical Society},
  doi = {10.1103/PhysRevX.10.021041},
  url = {https://link.aps.org/doi/10.1103/PhysRevX.10.021041}
}

@Article{Yang2020,
	Author = {Yang, Bing and Sun, Hui and Ott, Robert and Wang, Han-Yi and Zache, Torsten V. and Halimeh, Jad C. and Yuan, Zhen-Sheng and Hauke, Philipp and Pan, Jian-Wei},
	Doi = {10.1038/s41586-020-2910-8},
	Isbn = {1476-4687},
	Journal = {Nature},
	Number = {7834},
	Pages = {392--396},
	Title = {Observation of gauge invariance in a 71-site Bose--Hubbard quantum simulator},
	Url = {https://doi.org/10.1038/s41586-020-2910-8},
	Volume = {587},
	Year = {2020}
}

\end{document}